\documentclass[12pt, letterpaper]{article}
\usepackage[utf8]{inputenc}
\usepackage{graphicx}
\graphicspath{ {images/} }
\usepackage{balance}
\usepackage{comment}
\usepackage{indentfirst}
\usepackage{scrextend}
\title{\textbf{Detection of asteroid trails in Hubble Space Telescope images using Deep Learning}}
\author{ANDREI A. PARFENI\footnote{International Computer High School of Bucharest, Romania}\\LAUREN\c TIU I. CARAMETE\footnote{Space Science Institute, M\u agurele, Romania}\\ANDREEA M. DOBRE\footnote{International Computer High School of Constan\c ta, Romania}\\NGUYEN TRAN BACH\textsuperscript{*}\\}
\date{October 2020}

\begin{document}

\maketitle

\begin{center}
\large
\textbf{ABSTRACT}
\end{center}
\indent{
We present an application of Deep Learning for the image recognition of asteroid trails in single-exposure photos taken by the Hubble Space Telescope. Using algorithms based on multi-layered deep Convolutional Neural Networks, we report accuracies of above 80\% on the validation set. Our project was motivated by the Hubble Asteroid Hunter project on Zooniverse, which focused on identifying these objects in order to localize and better characterize them. We aim to demonstrate that Machine Learning techniques can be very useful in trying to solve problems that are closely related to Astronomy and Astrophysics, but that they are still not developed enough for very specific tasks.}

\tableofcontents

\section{Introduction}

\indent{
Machine Learning (ML) is defined as the study of algorithms that improve their performances over given tasks when trained using input data. While the field itself has existed for more than half a century, it is only in the last 15 years that ML algorithms have begun to garner more attention from researchers and companies that do work in other fields, as their performances (measured by both the accuracies, and the real time necessary to train and then to implement the algorithms) have improved enough to become relevant and useful for many tasks \cite{qiu2016survey}.

While many subfields of ML, such as Natural Language Processing, autonomous driving and speech synthesis, have seen important breakthroughs in the last few years, it is the study of image classification that has been most impacted by the development of Deep Learning, more specifically the realization that models called Convolutional Neural Networks allowed for greater accuracies and for lower costs in labeling the data than ever before \cite{yamashita2018convolutional}. Combined with the increasing data set sizes and the creation of better CPUs that allowed for faster computation \cite{lopes2014towards}, Deep Learning algorithms have become the mainstay of top-level ML research in this subject.

Therefore, in recent times, neural networks have been one of the most useful tools for astronomers in trying to recognize and classify many types of objects in our solar system, as the field has become much more data-rich. For instance, radio recordings have been successfully used to detect and extract samples of meteors entering the Earth's atmosphere \cite{roman2014automatic}. Direct image recognition has also been successful in identifying stars, galaxies and quasars \cite{martinazzo2020self}. Because comets and asteroids, together named Near Earth Objects, can pose a serious threat to our planet \cite{atkinson2001risks}, while also being useful case studies for the development of our understanding of the solar system, it is important that we apply the same proven methods to automatically detect asteroid trails.}

\section{Data set}

Our data set was made up of 2000 images, out of which 1036 contained asteroid trails, while 964 did not. While some of these photos were obtained from the Zooniverse site itself, most of them were offered by Sandor Kruk, researcher working at ESA.

\begin{figure}[h]
    \centering
    \includegraphics[width=0.4\textwidth]{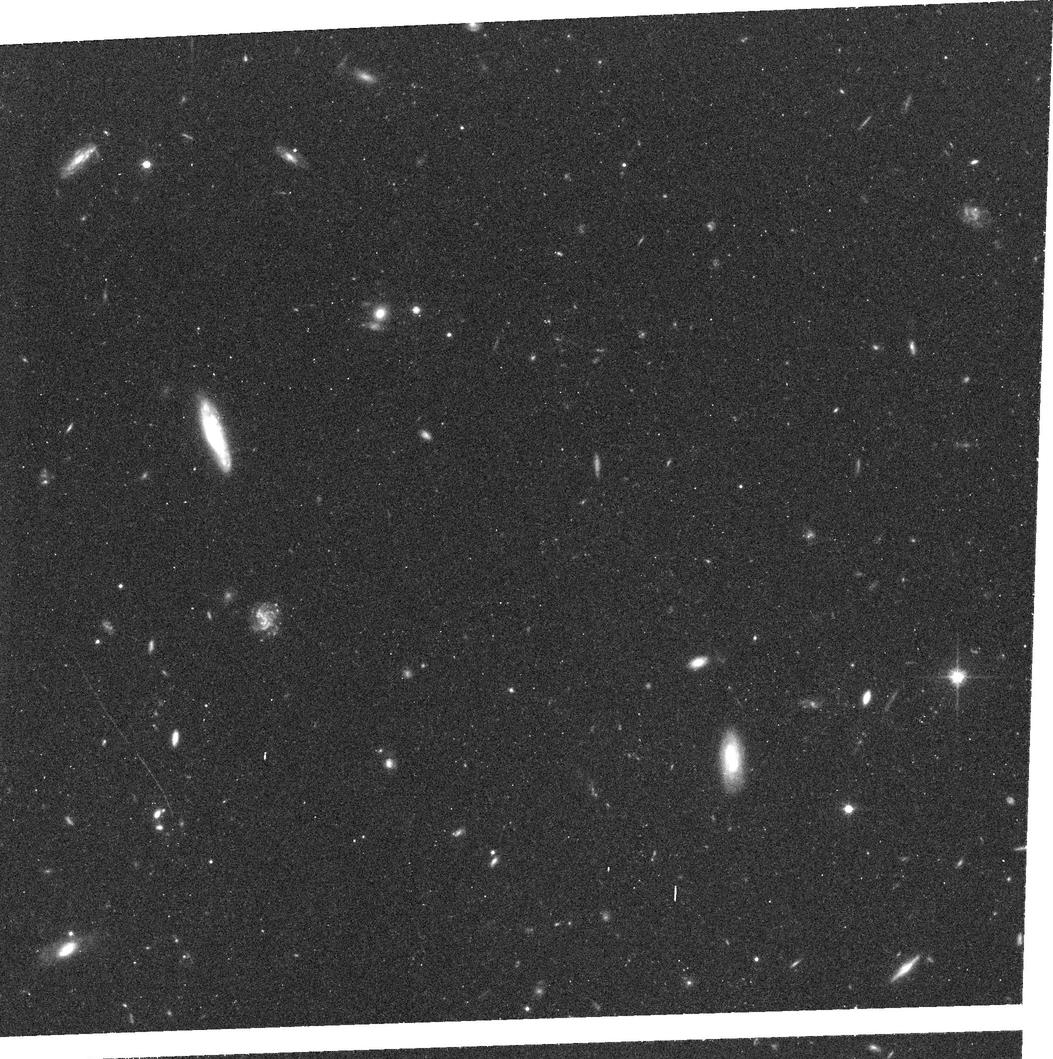}
    \caption{Asteroid trails}
    \label{fig:example_asteroid}
\end{figure}

\begin{figure}[ht]
    \centering
    \includegraphics[width=0.4\textwidth]{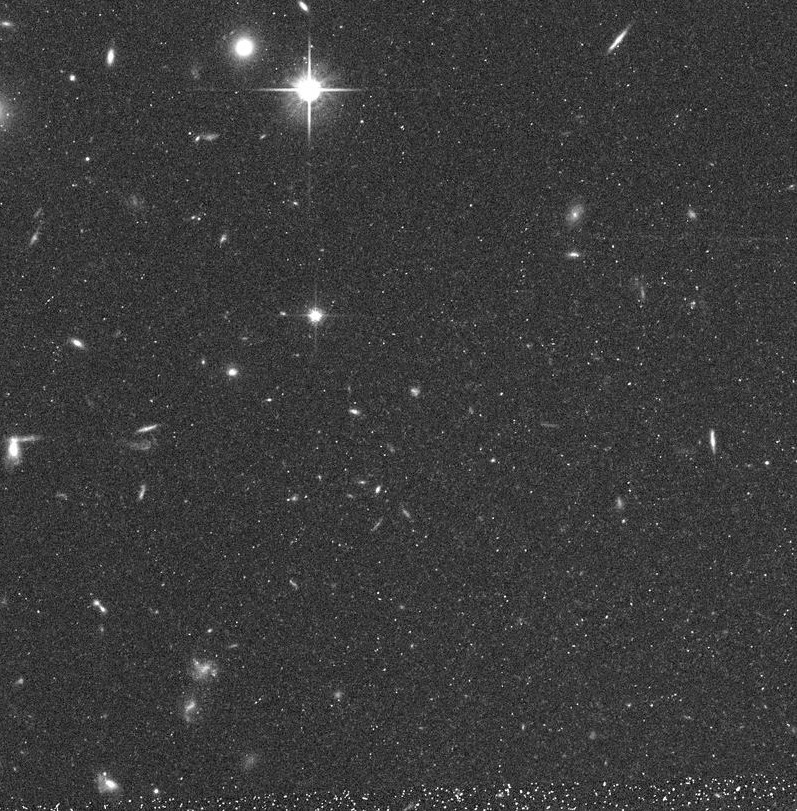}
    \caption{No asteroid trails}
    \label{fig:example_no_asteroid}
\end{figure}

The sizes of the unprocessed images varied. Many of them had sizes between 1040-1060x1050-1070 px, while others had sizes ranging from 400-800x400-800 px. These images were all almost square (with small differences between their lengths and widths). An example of an image containing an asteroid trail (on the bottom-left side) is provided in Figure \ref{fig:example_asteroid}, with a size of 1053x1059 px. Another example, of an image containing no asteroid trails, is provided in Figure \ref{fig:example_no_asteroid}, with a size of 797x811 px.

As seen in Figure \ref{fig:example_asteroid}, some of the pictures taken were imperfect, as the process of taking them was not uniform. However, the presence of the white pixels and the fact that the images are not perfect rectangles were factors that could be mitigated easily by reshaping the images and normalizing the pixel values.

\section{Methodology}

The sequence of steps taken in order to train and evaluate the models consisted of splitting the images into training and validation sets, preprocessing and augmenting the dataset, building and training multiple Convolutional Neural Networks in Python, selecting the ones that performed best on the training set and evaluating them on the validation set.

\subsection{The data split}
We decided to split the data into two separate sets (training and validation), using the recommended 80\%-20\% split.

This statistical split guaranteed us that the model would be properly trained on a sufficiently large set of images, while also avoiding any possible over fitting problems, thus resulting in accuracies that would closely match its performance on new, real-life data.

\subsection{Data preprocessing}

We used the ImageDataGenerator (IDG) class from Keras (running on top of TensorFlow, in the programming language Python) in order to preprocess the data set.

We utilized it to reshape the images to the target size of 256x256 px and to change them to a grayscale representation, which allowed for fast training of the programs, while maintaining the relevant features of the photos. We also normalized the image matrices by rescaling the values of the pixels to be in the range [0, 1].

The ImageDataGenerator class created a batch of 32 augmented images for every image that was introduced, by randomly rotating, zooming, shifting and shearing the image (all within small, carefully-constructed ranges), thus artificially outputting a larger training set of 51200 images that we could train the model on.

\begin{figure}[ht]
    \centering
    \includegraphics[width = 0.4\textwidth]{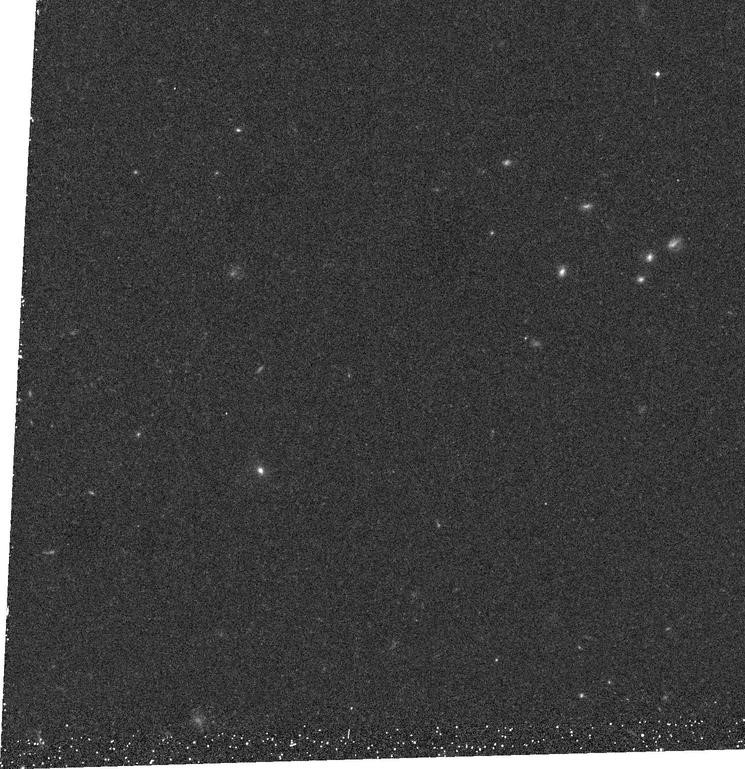}
    \caption{Before using the IDG class}
    \label{fig:example_initial_IDG}
\end{figure}

\begin{figure}[!tbp]
    \centering
    \begin{minipage}[b]{0.3\textwidth}
        \includegraphics[width = \textwidth]{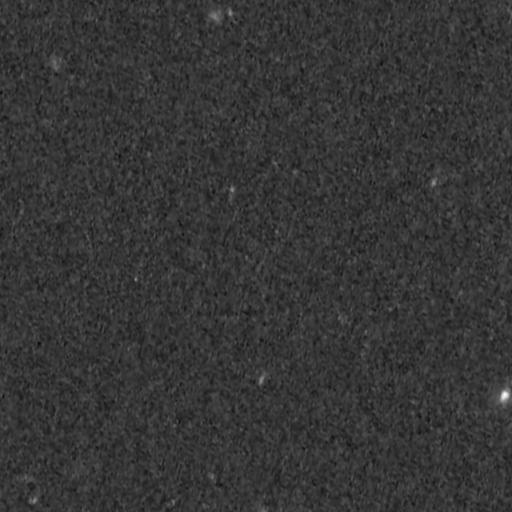}
    \end{minipage}
    \begin{minipage}[b]{0.3\textwidth}
        \includegraphics[width = \textwidth]{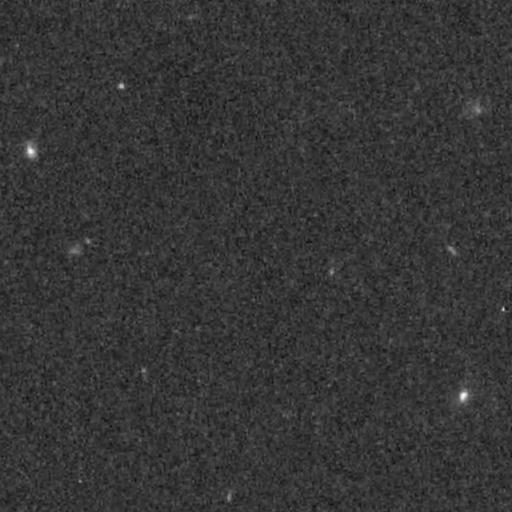}
    \end{minipage}
    \begin{minipage}[b]{0.3\textwidth}
        \includegraphics[width = \textwidth]{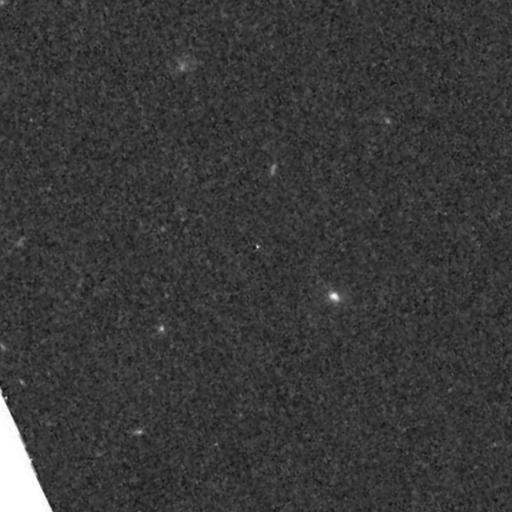}
    \end{minipage}
    \begin{minipage}[b]{0.3\textwidth}
        \includegraphics[width = \textwidth]{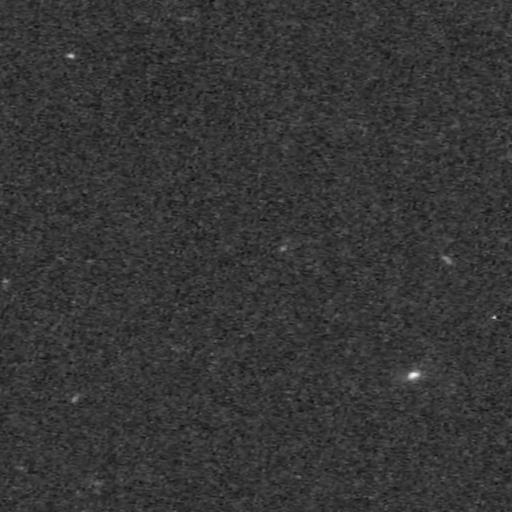}
    \end{minipage}
    \begin{minipage}[b]{0.3\textwidth}
        \includegraphics[width = \textwidth]{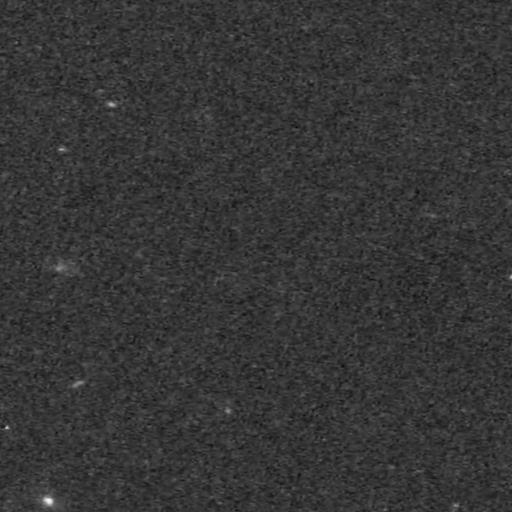}
    \end{minipage}
    \caption{After using the IDG class}
    \label{fig:example_after_IDG}
\end{figure}

Figure \ref{fig:example_initial_IDG} shows an example of a photo not containing asteroid trails, taken from the Zooniverse site.

Meanwhile, Figure \ref{fig:example_after_IDG} shows an example of what 5 of the 32 artificially generated photos could look like. All these photos were created from the image displayed in Figure \ref{fig:example_initial_IDG} by using transformations similar to the ones described above. These generated photos were all zoomed in to some extent, and thus they all mostly showed relatively different parts of the initial image, however the actual transformations used smaller zooms than these, ensuring that no asteroid trails were accidentally removed from the asteroid trail-containing training set.

\subsection{Artificial Neural Networks}

\begin{figure}[ht]
    \centering
    \includegraphics[width = 0.9\textwidth]{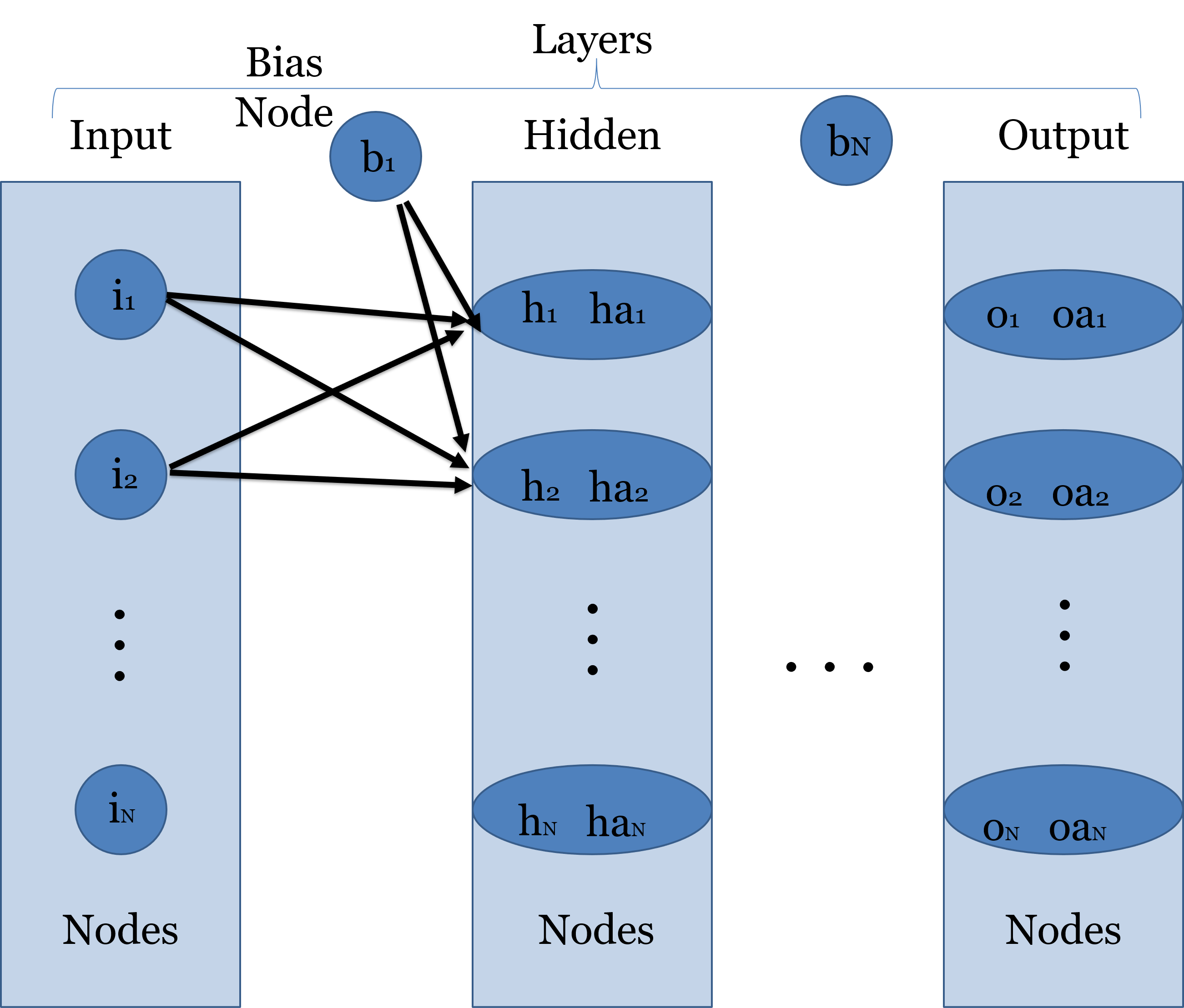}
    \caption{Standard structure of a feedforward ANN}
    \label{fig:example_multilayer_perceptron}
\end{figure}

The models we employed built upon the standard Artificial (Feedforward) Neural Network (ANN) architecture used by Multilayer Perceptrons (shown in Figure \ref{fig:example_multilayer_perceptron}), which consist of an input layer, followed by one or more hidden layers that are tied together, one after another, for which the numbers in the nodes are calculated as the values of the activation function of a linear combination between the weights and bias terms and the values of the preceding layer, and by the output layer, whose activation function is typically either the logistic function, or the rectified linear unit. The weights inherent to these models are typically trained through an optimized version of the process of gradient descent, generally  using the backpropagation algorithm.

\begin{figure}[ht]
    \centering
    \includegraphics[width = 0.9\textwidth]{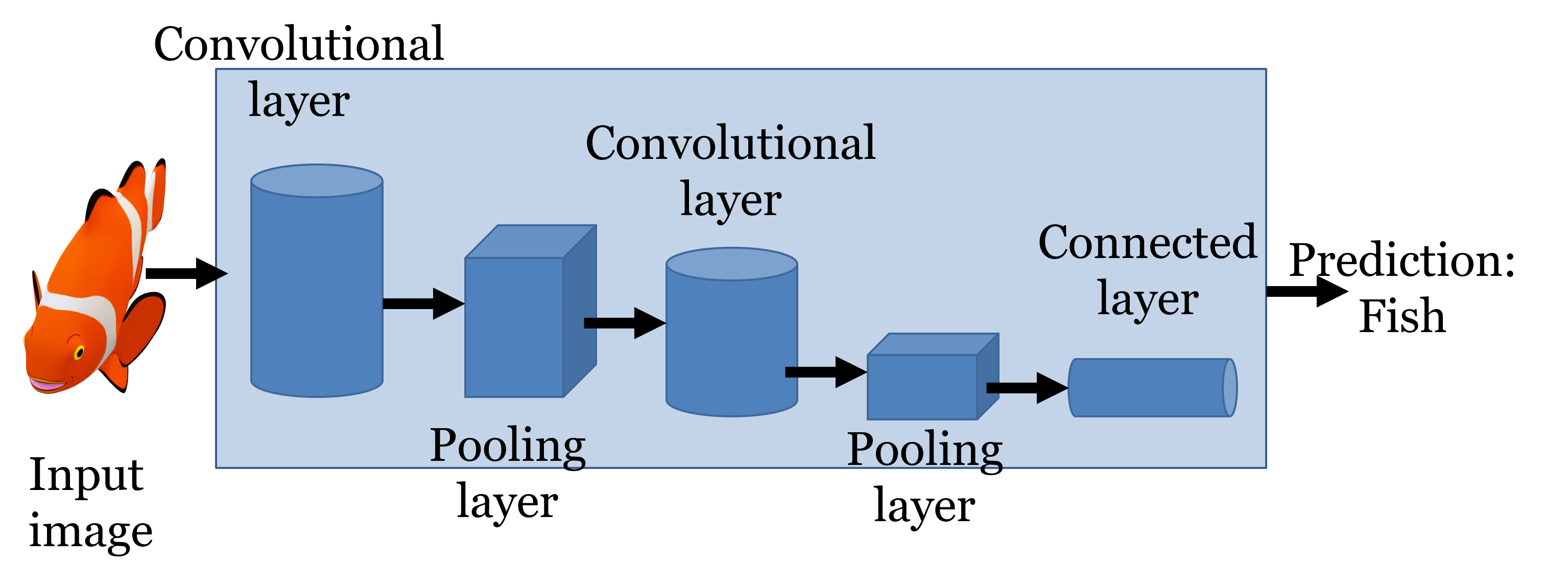}
    \caption{Standard structure of a CNN}
    \label{fig:example_convolutional_ANN}
\end{figure}

However, the models that performed best on the validation and test sets were Convolutional Neural Networks (CNNs), versions of feedforward ANNs that employed two pairs of Convolutional and Max Pooling layers. The Convolutional layers first performed an operation called convolution on the initial 256x256 array of pixels corresponding to the image, which consisted of creating multiple new two-dimensional arrays through the dot product between iterated sections of the image and carefully-selected 2x2 filters, thus capturing the crucial features embedded in the pixel array. Then, the models employed the Pooling layers, a form of dimensionality reduction which selects only the maximum pixel value in each section of the image, thus making the model more robust and cheaper to train (other versions of Pooling layers also exist, such as Average Pooling, but Max Pooling is typically considered most suitable for these types of problems). The models were completed by the addition of one deep layer and the final output one, which used a sigmoid activation function. In order to reduce overfitting problems, the models also employed two Dropout layers.

Figure \ref{fig:example_convolutional_ANN} shows an example of what a CNN architecture looks like. The increased use of Convolutional layers has been the key for the development of Deep Learning as a very useful means of recognition and prediction \cite{goodfellow2016deep}, especially for tasks involving computer vision \cite{goodfellow2013multi}, so it is no surprise that models based upon it perform better than the general-use algorithms do.

\subsection{Training and Evaluation}
The process of training and evaluating the algorithms was done using of Google Colab, a browser-based notebook environment similar to Jupyter Notebook. This was done in order to benefit from the free availability of Graphical Processing Units and Tensor Processing Units, which sped up the learning process significantly.

The optimizer we used is called Stochastic Gradient Descent, and the models were trained for 500 epochs at the learning rate of 0.001. The models aimed to minimize the Binary Crossentropy loss function on the training set.

Because the data sets were balanced, the models were compared based on the accuracies they displayed on the validation sets.

\section{Results and Discussion}

\begin{figure}[ht]
    \centering
    \includegraphics[width = 0.9\textwidth]{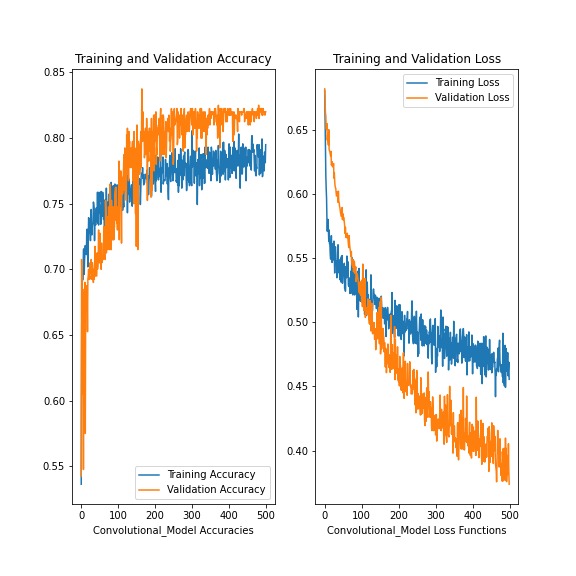}
    \caption{Plot of the model's performance over time}
    \label{fig:results}
\end{figure}

The best models obtained an accuracy of 82\% on the validation set, with a Binary Crossentropy loss function of 0.3893. Figure \ref{fig:results} shows a plotting of the accuracies and loss functions that the model obtained over the 500 epochs.

\begin{figure} [!tbp]
    \centering
    \begin{minipage}[b]{0.45\textwidth}
        \includegraphics[width = \textwidth]{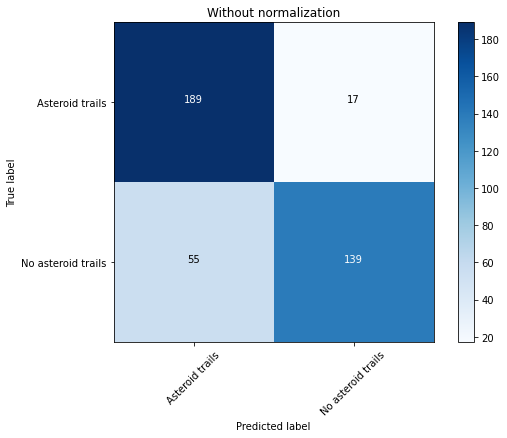}
    \end{minipage}
    \hfill
    \begin{minipage}[b]{0.45\textwidth}
        \includegraphics[width = \textwidth]{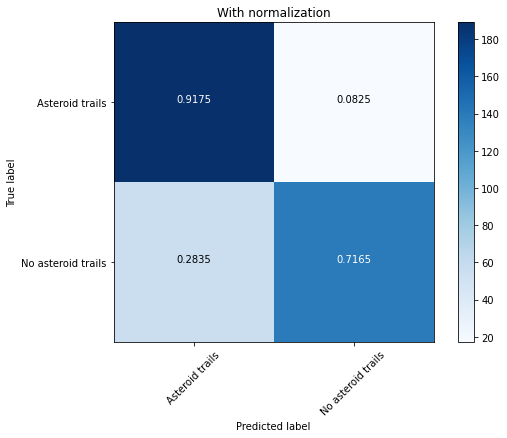}
    \end{minipage}
    \caption{Confusion matrices on the validation set}
    \label{fig:confusion_matrices}
\end{figure}

As can be seen from the confusion matrices in Figure \ref{fig:confusion_matrices}, the precision of the predictions is 77.45\%, while the recall is 91.74\%, for an overall F1 score of 0.84.

Since the neural network's accuracies on the validation set were similar to and in some cases even exceeded those on the training set, there appears to be almost no sign of over fitting, which means that carefully increasing its depth would possibly result in even better results. However, this must be balanced out with the fact that a large model would contain more than 40 million parameters and would thus be very hard to properly train. It also means that increasing the size of the training data set, whether through the capture of more images or through methods of synthetic data generation, would probably not have improved the accuracy as much.

While the overall performance of the model was good, the relatively high difference between its recall and its precision made the F1 score greater than the accuracy, which showed that the model predicts that images contain asteroid trails far too often.

The biggest problem the model faced was the relative subtlety of the asteroid trails in the images. They were mostly straight lines of a small width, and such simple shapes were not as easily recognized by deep Convolutional Neural Networks as more complex ones. On the other hand, one of their built-in advantages was the ability to add augmented data in the training sets with minimal additional computational cost, which is not the case with other classifiers, such as Support Vector Machines.

\section{Conclusion}

Based on the resulting data, Deep Learning models using convolutional layers, trained on an artificially generated training set of 51200 images, performed adequately in classifying images of asteroid trails in space photos, identifying them with accuracies of above 80\% on the validation set.

There are several ideas and techniques which could improve the performance of future models on similar tasks. For instance, utilizing large batches of images of the same size and of a square shape could help avoid any reshape-induced problems. Properly identifying and removing the large stars present in the photos could eliminate much of the noise in the images and thus allow the algorithms to learn how to identify asteroid trails faster. Using varying dilation rates for the convolutional kernels and implementing Average Pooling layers instead of Max Pooling ones should also be considered.

Different architectures or types of models could also provide a starting point for avoiding the issues that the CNNs faced. Combining the neural network architecture with the kernel method, more commonly seen in Support Vector Machines, has recently achieved outstanding results on the CIFAR-10 dataset \cite{shankar2020neural}, and similar models could benefit from combining the advantages of SVM-like methods (such as identifying simpler shapes) with those of Artificial Neural Networks.

The continued development of Machine Learning algorithms bodes well for the fields of Astronomy and Astrophysics. While ML remains under utilized at the moment, models such as ours show that Deep Learning has great applicability in these areas.

\bibliography{bibliography}

\begin{thebibliography}{1}

\bibitem{qiu2016survey}
Junfei Qiu, Qihui Wu, Guoru Ding, Yuhua Xu, and Shuo Feng.
\newblock A survey of machine learning for big data processing.
\newblock {\em EURASIP Journal on Advances in Signal Processing}, 2016(1):67,
  2016.

\bibitem{yamashita2018convolutional}
Rikiya Yamashita, Mizuho Nishio, Richard Kinh~Gian Do, and Kaori Togashi.
\newblock Convolutional neural networks: an overview and application in
  radiology.
\newblock {\em Insights into imaging}, 9(4):611--629, 2018.

\bibitem{lopes2014towards}
Noel Lopes and Bernardete Ribeiro.
\newblock Towards adaptive learning with improved convergence of deep belief
  networks on graphics processing units.
\newblock {\em Pattern recognition}, 47(1):114--127, 2014.

\bibitem{roman2014automatic}
VS~Roman and C{\u{a}}t{\u{a}}lin Buiu.
\newblock Automatic detection of meteors using artificial neural networks.
\newblock In {\em Proceedings of the International Meteor Conference, Giron,
  France}, pages 18--21, 2014.

\bibitem{martinazzo2020self}
Ana Martinazzo, Mateus Espadoto, and Nina~ST Hirata.
\newblock Self-supervised learning for astronomical image classification.
\newblock {\em arXiv preprint arXiv:2004.11336}, 2020.

\bibitem{atkinson2001risks}
Harry Atkinson.
\newblock Risks to the earth from impacts of asteroids and comets.
\newblock {\em europhysics news}, 32(4):126--129, 2001.

\bibitem{goodfellow2016deep}
Ian Goodfellow, Yoshua Bengio, Aaron Courville, and Yoshua Bengio.
\newblock {\em Deep learning}, volume~1.
\newblock MIT press Cambridge, 2016.

\bibitem{goodfellow2013multi}
Ian~J Goodfellow, Yaroslav Bulatov, Julian Ibarz, Sacha Arnoud, and Vinay Shet.
\newblock Multi-digit number recognition from street view imagery using deep
  convolutional neural networks.
\newblock {\em arXiv preprint arXiv:1312.6082}, 2013.

\bibitem{shankar2020neural}
Vaishaal Shankar, Alex Fang, Wenshuo Guo, Sara Fridovich-Keil, Ludwig Schmidt,
  Jonathan Ragan-Kelley, and Benjamin Recht.
\newblock Neural kernels without tangents.
\newblock {\em arXiv preprint arXiv:2003.02237}, 2020.

\end{thebibliography}
\end{document}